\documentstyle[12pt,fullpage]{article}
\begin{document}

\begin{center}
{\Large \bf
Formation of Root Singularities on the Free Surface\\
of a Conducting Fluid in an Electric Field}

\bigskip

{\large\bf N. M. Zubarev}

\bigskip
{\small\sl Institute of Electrophysics, Ural Division, Russian Academy of
Sciences,\\
106 Amundsena Street, 620016 Ekaterinburg, Russia\\
e-mail: nick@ami.uran.ru}
\end{center}

\begin{abstract}
{\small
The formation of singularities on a free surface of a conducting ideal
fluid in a strong electric field is considered. It is found that the
nonlinear equations of two-dimensional fluid motion can be solved in the
small-angle approximation. This enables us to show that for almost
arbitrary initial conditions the surface curvature becomes infinite in a
finite time.
}
\end{abstract}

\bigskip

Electrohydrodynamic instability of a free surface of a conducting
fluid in an external electric field [1,2] plays an essential role in
a general problem of the electric strength. The interaction of
strong electric field with induced charges at the surface of the fluid
(liquid metal for applications) leads to the avalanche-like growth of
surface perturbations and, as a consequence, to the formation of
regions with high energy concentration which destruction can be
accompanied by intensive emissive processes.

In this Letter we will show that the nonlinear equations of motion of a
conducting fluid can be effectively solved in the approximation of small
perturbations of the boundary. This allows us to study the nonlinear
dynamics of the electrohydrodynamic instability and, in particular, the
most physically meaningful singular solutions.

Let us consider an irrotational motion of a conducting ideal fluid with a
free surface, $z=\eta(x,y,t)$, that occupies the region
$-\infty<z\leq\eta(x,y,t)$, in an external uniform electric field $E$.
We will assume the influence of gravitational and capillary
forces to be negligibly small, which corresponds to the condition
$$
E^2\gg 8\pi\sqrt{g\alpha\rho},
$$
where $g$ is the acceleration of gravity, $\alpha$ is the surface tension
coefficient, and $\rho$ is the mass density.

The potential of the electric field $\varphi$ satisfies the Laplace
equation,
$$
\Delta\varphi=0,
$$
with the following boundary conditions,
$$
\varphi\to -Ez, \qquad z\to\infty,
$$
$$
\varphi=0, \qquad z=\eta.
$$
The velocity potential $\Phi$ satisfies the incompressibility equation
$$
\Delta\Phi=0,
$$
which one should solve together with the dynamic and kinematic relations on
the free surface,
$$
\frac{\partial\Phi}{\partial t}+\frac{(\nabla\Phi)^2}{2}=
\frac{(\nabla\varphi)^2}{8\pi\rho}+F(t), \qquad z=\eta,
$$
$$
\frac{\partial\eta}{\partial t}=\frac{\partial\Phi}{\partial z}
-\nabla\eta\cdot\nabla\Phi,
\qquad z=\eta,
$$
where $F$ is some function of variable $t$, and the boundary condition
$$
\Phi\to 0, \qquad z\to-\infty.
$$
The quantities $\eta(x,y,t)$ É $\psi(x,y,t)=\Phi|_{z=\eta}$ are
canonically conjugated, so that the equations of motion take the
Hamiltonian form [3],
$$
\frac{\partial\psi}{\partial t}=-\frac{\delta H}{\delta\eta},
\qquad
\frac{\partial\eta}{\partial t}=\frac{\delta H}{\delta\psi},
$$
where the Hamiltonian
$$
H=\int\limits_{z\leq\eta}\frac{(\nabla\Phi)^2}{2} d^3 r
-\int\limits_{z\geq\eta}\frac{(\nabla\varphi)^2}{8\pi\rho} d^3 r
$$
coincides with the total energy of a system. With the help of the Green
formula it can be rewritten as the surface integral,
$$
H=\int\limits_{s}\left[\frac{\psi}{2}\,\frac{\partial\Phi}{\partial n}+
\frac{E\eta}{8\pi\rho}\,\frac{\partial\tilde\varphi}{\partial n}\right]ds,
$$
where $\tilde\varphi=\varphi+Ez$ is the perturbation of the electric field
potential; $ds$ is the surface differential.

Let us assume $|\nabla\eta|\ll 1$, which corresponds to the approximation
of small surface angles. In such a case we can expand the integrand in a
power series of canonical variables $\eta$ and $\psi$. Restricting
ourselves to quadratic and cubic terms we find after scale transformations
$$
t\to t E^{-1}(4\pi\rho)^{1/2},
\quad
\psi\to\psi E/(4\pi\rho)^{1/2},
\quad
H\to HE^2/(4\pi\rho)
$$
the following expression for the Hamiltonian,
$$
H=\frac{1}{2}\int\left[\psi\hat k\psi+
\eta\left((\nabla\psi)^2-(\hat k\psi)^2\right)\right] d^2 r
$$
$$
-\frac{1}{2}\int\left[\eta\hat k\eta-\eta\left((\nabla\eta)^2-
(\hat k\eta)^2\right)\right] d^2 r.
$$
Here $\hat k$ is the integral operator with the difference kernel, whose
Fourier transform is the modulus of the wave vector,
$$
\hat{k}f=-\frac{1}{2\pi}\!\int\limits_{-\infty}^{+\infty}
\int\limits_{-\infty}^{+\infty}
\frac{f(x',y')}{\left[(x'-x)^2+(y'-y)^2\right]^{3/2}}\,dx'dy'.
$$
The equations of motion, corresponding to this Hamiltonian, take the
following form,
\begin{equation}
\psi_t-\hat k\eta=\frac{1}{2}\left[(\hat k\psi)^2-(\nabla\psi)^2+
(\hat k\eta)^2-(\nabla\eta)^2\right]+
\hat k(\eta\hat k\eta)+\nabla(\eta\nabla\eta),
\end{equation}
\begin{equation}
\eta_t-\hat k\psi=-\hat k(\eta\hat k\psi)-\nabla(\eta\nabla\psi).
\end{equation}
Subtraction of Eqs.~(2) and (1) gives in the linear approximation
the relaxation equation
$$
(\psi-\eta)_t=-\hat k(\psi-\eta),
$$
whence it follows that we can set $\psi=\eta$ in the nonlinear terms of
Eqs.~(1) and (2), which allows us to simplify the equations of motion.
Actually, adding Eqs.~(1) and (2) we obtain an equation
for a new function $f=(\psi+\eta)/2$,
\begin{equation}
f_t-\hat k f=\frac{1}{2}\,(\hat k f)^2-\frac{1}{2}\,(\nabla f)^2,
\end{equation}
which corresponds to the consideration of the growing branch of the
solutions. As $f=\eta$ in the linear approximation, Eq.~(3) governs
the behavior of the elevation $\eta$.

First we consider the one-dimensional case when function $f$
depends only on $x$ (and $t$) and the integral operator $\hat k$ can be
expressed in terms of the Hilbert transform $\hat H$,
$$
\hat k=-\frac{\partial}{\partial x}\,\hat H,
\qquad
\hat{H}f=\frac{1}{\pi}\,\mbox{P}\!\!\int\limits_{-\infty}^{+\infty}
\frac{f(x')}{x'-x}\,dx',
$$
where P denotes the principal value of the integral. As a result, Eq.~(3)
can be rewritten as
\begin{equation}
f_t+\hat H f_x=\frac{1}{2}\,(\hat H f_x)^2-\frac{1}{2}\,(f_x)^2.
\end{equation}
It should be noted that if one introduces a new function
$\tilde f=\hat H f$, then Eq.~(4) transforms into the equation
proposed in Ref. [4] for the description of the nonlinear stages of the
Kelvin-Helmholtz instability.

For further consideration it is convenient to introduce a function,
analytically extendable into the upper half-plane of the complex variable
$x$,
$$
v=\frac{1}{2}\,(1-i\hat H)f_x.
$$
Then Eq.~(4) takes the form
$$
\mbox{Re}\left(v_t+iv_x+2vv_x\right)=0,
$$
that is, the investigation of integro-differential equation (4)
amounts to the analysis of the partial differential equation
\begin{equation}
v_t+iv_x+2vv_x=0,
\end{equation}
which describes the wave breaking in the complex plane. Let us study this
process in analogy with [5,6], where a similar problem was considered.
Eq. (5) can be solved by the standard method of characteristics,
\begin{equation}
v=Q(x'),
\end{equation}
\begin{equation}
x=x'+it+2Q(x')t.
\end{equation}
where the function $Q$ is defined from initial conditions. It is clear
that in order to obtain an explicit form of the solution we must resolve
Eq.~(7) with respect to $x'$. A mapping $x\to x'$, defined by Eq.~(7),
will be ambiguous if $\partial x/\partial x'=0$ in some point, i.e.
\begin{equation}
1+2Q_{x'}t=0.
\end{equation}
Solution of (8) gives a trajectory $x'=x'(t)$ on the complex plane $x'$.
Then the motion of the branch points of the function $v$ is defined
by an expression
$$
x(t)=x'(t)+it+2Q(x'(t))t.
$$
At some moment $t_0$ when the branch point touches the real axis, the
analiticity of $v(x,t)$ at the upper half-plane of variable $x$ breaks,
and a singularity appears in the solution of Eq.~(4).

Let us consider the solution behavior close to the singularity. Expansion
of (6) and (7) at a small vicinity of $x=x(t_0)$ up to the leading orders
gives
$$
v=Q_0-\delta x'/(2t_0),
$$
$$
\delta x=i\delta t+2Q_0\delta t+Q''t_0(\delta x')^2,
$$
where $Q_0=Q(x'(t_0))$, $Q''=Q_{x'x'}(x'(t_0))$,
$\delta x\!=\!x\!-\!x(t_0)$,
$\delta x'\!=\!x'\!-\!x'(t_0)$, and
$\delta t\!=\!t\!-\!t_0$.
Eliminating $\delta x'$ from these equations, we find that
close to singularity $v_x$ can be represented in the self-similar form
($\delta x\sim\delta t$),
$$
v_x=-\left[16Q''t_0^3
(\delta x-i\delta t-2Q_0\delta t)\right]^{-1/2}.
$$
As $\mbox{Re}(v)=\eta/2$ in the linear approximation, we have at $t=t_0$
$$
\eta_{xx}\sim|\delta x|^{-1/2},
$$
that is the surface curvature becomes infinite in a finite time. It should
be mentioned that such a behavior of the charged surface is similar to
the behavior of a free surface of an ideal fluid in the absence of
external forces [5,6], though the singularities are of a different nature
(in the latter case the singularity formation is connected with inertial
forces).

Let us show that the solutions corresponding to the root singularity
regime are consistent with the applicability condition of the truncated
equation (3). Let $Q(x')$ be a rational function with one pole in the lower
half-plane,
\begin{equation}
Q(x')=-\frac{is}{2(x'+iA)^2},
\end{equation}
which corresponds to the spatially localized one-dimensional perturbation of
the surface ($s>0$ and $A>0$). The characteristic surface angles are
thought to be small, $\gamma\approx s/A^2\ll 1$.

It is clear from the symmetries of (9) that the most rapid branch point
touches the real axis at $x=0$. Then the critical moment $t_0$ can be
found directly from Eqs.~(7) and (8). Expansion of $t_0$ with respect to
the small parameter $\gamma$ gives
\begin{equation}
t_0\approx A\left[1-3(\gamma/4)^{1/3}\right].
\end{equation}
Taking into account that the evolution of the surface perturbation can be
described by an approximate formula
$$
\eta(x,t)=\frac{s(A-t)}{(A-t)^2+x^2},
$$
we have for the dynamics of the characteristic angles
$$
\gamma(t)\approx\frac{s}{(A-t)^2}.
$$
Then, substituting the expression for $t_0$ (10) into this
formula, we find that at the moment of the singularity formation with
the required accuracy
$$
\gamma(t_0)\sim\gamma^{1/3},
$$
that is, the angles remain small and the root singularities are consistent
with our assumption about small surface angles.

In conclusion, we would like to consider the more general case where the
weak dependence of all quantities from the spatial variable
$y$ is taken into account. One can find that if the condition
$|k_x|\ll|k_y|$ holds for the characteristic wave numbers, then the
evolution of the fluid surface is described by an equation
$$
\left[v_t+iv_x+2vv_x\right]_x=-iv_{yy}/2,
$$
which extends Eq.~(5) to the two-dimensional case.

An interesting group of particular solutions of this equation can be
found with the help of substitution $v(x,y,t)=w(z,t)$, where
$$
z=x-\frac{i}{2}\,\frac{(y-y_0)^2}{t}.
$$
The equation for $w$ looks like
$$
w_t+iw_z+2ww_z=-w/(2t).
$$
It is integrable by the method of characteristics, so that we can study
the analyticity violation similarly to the one-dimensional case.
Considering a motion of branch points in the complex plane of the
variable $z$ we find that a singularity arises at some moment $t_0<0$ at
the point $y_0$ along the $y$-axis. Close to the singular point at the
critical moment $t=t_0$ we get
$$
\left.\eta_{xx}\right|_{\delta y=0}\sim|\delta x|^{-1/2},
\qquad
\left.\eta_{xx}\right|_{\delta x=0}\sim|\delta y|^{-1}.
$$
This means that in the examined quasi-two-dimensional case the
second derivative of the surface profile becomes infinite at a single
isolated point.

Thus, the consideration of the behavior of a conducting fluid surface
in a strong electric field shows that the nonlinearity determines the
tendency for the formation of singularities of the root character,
corresponding to the surface points with infinite curvature. We can assume
that such weak singularities serve as the origin of the more powerful
singularities observed in the experiments [7,8].
\bigskip

I would like to thank A.M.~Iskoldsky and N.B.~Volkov for helpful
discussions, and E.A. Kuznetsov for attracting my attention to
Refs.~[5,6]. This work was supported by Russian Foundation for Basic
Research, Grant No.~97--02--16177.

\bigskip
\begin{center}
{\large \bf References}
\end{center}

\begin{enumerate}
\item L. Tonks, Phys. Rev. 48 (1935) 562.
\item Ya.I. Frenkel, Zh. Teh. Fiz. 6 (1936) 347.
\item V.E. Zakharov, J. Appl. Mech. Tech. Phys. 2 (1968) 190.
\item S.K. Zhdanov and B.A. Trubnikov, Sov. Phys. JETP 67 (1988) 1575.
\item E.A. Kuznetsov, M.D. Spector, and V.E. Zakharov, Phys. Lett. A
182 (1993) 387.
\item E.A. Kuznetsov, M.D. Spector, and V.E. Zakharov, Phys. Rev. E
49 (1994) 1283.
\item M.D. Gabovich and V.Ya. Poritsky, JETP Lett. 33, (1981) 304.
\item A.V. Batrakov, S.A. Popov, and D.I. Proskurovsky, Tech. Phys. Lett.
19 (1993) 627.
\end{enumerate}

\end {document}